\documentclass[12pt,,letterpaper]{JHEP3}
\usepackage{graphicx}
\usepackage{subfigure}
\usepackage{epsfig}
\usepackage{slashed}
\usepackage{amsmath}
\usepackage{bm}
\usepackage{float}
\usepackage{subfig}

\newcommand{\be}{\begin{equation}}
\newcommand{\ee}{\end{equation}}
\newcommand{\bea}{\begin{eqnarray}}
	\newcommand{\eea}{\end{eqnarray}}

\title{Scrambling time for analogue black holes embedded in AdS space}
\author{ Qing-Bing Wang$^{~1}$, Ming-Hui Yu$^{~1}$, Xian-Hui Ge$^{~1}$\footnote{ Corresponding author. gexh@shu.edu.cn}\\
	$^{1}$Department of Physics, College of Sciences, Shanghai University, 200444 Shanghai, China\\
}

\abstract{ We propose a gedanken experiment on realizing thermofield double state (TFD) by using analog black holes and provide an approach to test the scrambling time. Through this approach, we demonstrate clearly how shock wave changes the TFD state as time evolves. As the whole system evolves forward in time, the perturbation of space-time geometry will increase exponentially. Finally, it will destroy the entanglement between the two states of the thermal field, and the mutual information between them is reduced to zero in the time scale of scrambling.
The results show that for perturbations of analogue black holes embedded in AdS space, the scale of the scrambling time is closely related to the logarithm of entropy of the black hole. The results provide further theoretical argument for the scrambling time, which can be further falsified in experiments.
}

\begin{document}
	
	\section{Introduction}

	Black holes are perhaps the most interesting and mysterious objects in the universe.  In recent years, there is some great progress in black hole physics, such as gravitational waves observed by LIGO and VIRGO \cite{Abbott2016,Abbott2017}, the image of black holes observed by the Event Horizon Telescope \cite{EHT2019}. These progresses not only inject new impetus into the development of physics, but also undoubtedly provide strong evidence that black holes are real astrophysical objects in the sky.
	However, considering the quantum aspects of black holes, one may need reply more on the tabletop experiments.
	
	There are a lot of experimental progresses in the study of simulated black holes or simulated gravity in recent years \cite{Nova,Cromb,Nation,Giovanazzi,Jacobson,Garay,Hu2019,Horstmann,Philbin}.
	In particular, the study of analogue black holes offers new viewpoints.
	In fact, as early as in 1981, Unruh proposed the possibility of simulating black holes by using fluid \cite{Unruh}. After that, physicists carried out observations of thermal Hawking radiation in cold atoms \cite{Nova} and completed the simulation of the Penrose process of black holes in the fluid experiment \cite{Cromb}.
	
	In addition, there are also many ways to mimic black holes or curved spacetime backgrounds \cite{Drori2019,Sheng2016,Sheng2013,Cheng2010}, such as lasers with optical metamaterials.
	For simulating black holes, Cheng et al. have realized an artificial ``electromagnetic black hole" in microwave band by using an artificial split-ring resonator (SRR) as a metamaterial \cite{Cheng2010}.
	Sheng et al. constructed waveguides with nonuniform refractive index according to centrosymmetric, and realized the equivalent potential field similar to the gravitational field of black holes \cite{Sheng2013}. The gravitational bending effect of black holes on light propagation can be well simulated.
	Koke et al. used waveguide arrays to simulate quantum field theory. By connecting the Dirac equation of curved space-time with coupled waveguide arrays, they studied the generation of positive and negative ion pairs in curved space-time and the Zitterbewegung effect (trembling motion) of general relativity \cite{Koke2016}.	
	Wang et al. studied the generation of positive and negative ion pairs near the event horizon of black holes \cite{Wang2020}.
	Tian et al. use the tunneling effect of trapped ions, which is analogous to Hawking radiation, to test the upper limit of the speed of scrambling \cite{Zehua}.
	
	For simulation of curved spacetime, related work proves that the nonlinear optics of hyperbolic metamaterials allow holographic dual description \cite{Smolyaninov2014}. The wave equation describing the propagation of light through metamaterials has a 2+1 Lorentz symmetry. The role of time in the corresponding 3D Minkowski spacetime is played by spatial coordinates aligned with the optical axis of the material.
	Other similar work \cite{Li1,Li2,Smolyaninov2010} proposed the simulation of the de Sitter (dS) spacetime and used metamaterials to simulate changes in the structure of space and time.

	On the other hand, most of the analogy and simulation studies about black holes or related involve quantum many body theory, in which entanglement plays an important role, and its quantification is called entanglement entropy. Considering a quantum many-body system with $N$ degrees of freedom in the pure state, for the subsystem with $m$ degrees of freedom less than $N/2$, the entanglement entropy of these subsystems will reach its maximum value when they are in thermal equilibrium, and the total system is called scrambled. One can simply understand that the information in the system is so evenly and thoroughly dissipated that we need to study at least $N/2$ degrees of freedom to capture it \cite{Sekino2008,Page1993}.	
	If we add a small amount of information or injection of a small amount of new degrees of freedom to the system, in the beginning, only the local degrees of freedom in the system are affected. One can restore the information with only  a small number of degrees of freedom. But in the end, the perturbation will spread to the whole system, the system returns to the scrambled state.
	In this process, the period of time from adding perturbation til the system returns to the equilibrium state again is called the scrambling time.

     The determination of scrambling time not only plays an important role in quantum information, but also has far-reaching significance in the black hole spacetime. According to the Hayden-Preskill experiment \cite{Hayden2007}, for a black hole has evaporated for a time bigger than half of its lifetime,
     the order of time required to reconstruct the information of the matter falling into the black hole is roughly the same as the scrambling time \cite{Sekino2008}. Therefore, here we are committed to providing some useful exploration for the experimental measurement of the scrambling time in the context of curved spacetime.

	In order to realize the holographic description of the above process in analog gravity, a pratical method is to embed acoustic black holes in Anti de-Sitter (AdS) spacetime. First of all, analogue black holes are easier to observe in experiments than astrophysical black holes.
	Secondly, it is convenient to calculate related physical quantities in AdS spacetime, such as the geodesic distance and the entanglement entropy. Thirdly, we find that when acoustic black holes are considered in AdS space-time, the geometry structure (explained in the next section of this paper) is similar to BTZ black holes. At the same time, it may also provide a new analogue for the experimental simulation of rotating black holes.

	For acoustic black holes embedded in the AdS space, we can obtain a metric expressed by the Hadamard product derived from the relativistic Gross-Pitaevskii theory or simply from the Yang-Mills theory \cite{GeXH2019}. Then, the (2+1)-dimensional acoustic black hole geometry of curved spacetime can be expressed as a product of the background metric and the acoustic metric \cite{GeXH2019,GeXH2020}
	\be\label{metric1}
	ds^2=(g^{GR}_{\mu\nu}\ast g^{ABH}_{\mu\nu})dx^{\mu}dx^{\nu}=
	\mathcal{G}_{tt}dt^2+\mathcal{G}_{rr}dr^2+\mathcal{G}_{\phi \phi}d\phi^2,
	\ee
	where $g^{GR}_{\mu\nu}$ is the background metric line-elements, and $g^{ABH}_{\mu\nu}$ is the acoustic metric line-elements.
	 Specifically in the fluid for the general acoustic metric \cite{Unruh,Visser}, we take
	\bea\label{metric}
	ds_{ABH}^2&=&-g_{tt}dt^2+g_{rr}dr^2+g_{\phi\phi}d\phi^2\nonumber\\&=&\!\frac{\rho_0}{c_s}\left[-(c_s^2-v^2)dt^2+\frac{c_s^2}{c_s^2-v^2}dr^2+r^2 d\phi^2\right]\!.
	\eea
	The density ratio of the fluid to the velocity of sound in the fluid $\frac{\rho_0}{c_s}$ can be absorbed to the left of the equation. Assumed that the flow velocity of the fluid $v$ goes in radially, in (2+1)-dimensional geometry, it can be set to $v=-\lambda/r$, where $\lambda$ is a parameter and the negative sign indicates that the fluid flows radially to the central region. The metric \eqref{metric} can be rewritten as
	\be
	ds_{ABH}^2=-c_s^2\left(1-\frac{\lambda^2}{c_s^2 r^2}\right)dt^2+\frac{1}{1-\frac{\lambda^2}{c_s^2 r^2}}dr^2+r^2 d\phi^2\!.
	\ee
	
	On the other hand, the (2+1)-dimensional Schwarschild-AdS metric is given by
	\be
	ds_{GR}^2=-r^2\left(1-\frac{r_0^2}{r^2}\right) dt^2+\frac{1}{r^2\left(1-\frac{r_0^2}{r^2}\right)} dr^2+r^2 d\phi ^2.
	\ee
	Notice that we take the AdS radius $\ell$ to be 1 and set $M=G=c=1$ for convenience. Considering the ratio between the sound velocity $c_s$ and the speed of light $c$, we take the value of $c_s$ to be $1/\sqrt 3$.
	Thus, we consider a (2+1)-dimensional acoustic black hole embedded in (2+1)-dimensional Schwarschild-AdS spacetime with the following line elements
	\bea
	\mathcal{G}_{tt}&=&-\frac{r^2}{3} \left(1-\frac{3\lambda^2}{r^2}\right) \left(1-\frac{r_0^2}{r^2}\right), \nonumber\\
	\mathcal{G}_{rr}&=&\frac{1}{r^2 \left(1-\frac{3\lambda^2}{r^2}\right) \left(1-\frac{r_0^2}{r^2}\right)} ,\label{factor} \\
	\mathcal{G}_{\phi \phi}&=&r^4.\nonumber
	\eea
	The acoustic event horizon $r_s$ locates at $r_s=\sqrt{3}\lambda$ and is larger than the event horizon $r_0$.
	For simplicity, we set $f(r)=\frac{r^2}{\sqrt{3}} \left(1-\frac{r_s^2}{r^2}\right) \left(1-\frac{r_0^2}{r^2}\right)$. In turn, the metric \eqref{metric1} can be recast as
	\be\label{metric2}
	ds^2=-f(r)dt^2+\frac{1}{f(r)}dr^2+\sqrt{3} r^4 d\phi ^2.
	\ee
	Compared \eqref{metric1} with \eqref{factor}, here we multiply a constant $\sqrt{3}$ on the last term.
	The acoustic Hawking temperature at the acoustic horizon is $T=f'(r_s)/(4\pi)=\frac{r_s^2-r_0^2}{2\sqrt{3}\pi r_s}$.

\begin{figure}[htbp]
		\centering
		\includegraphics[scale=0.4]{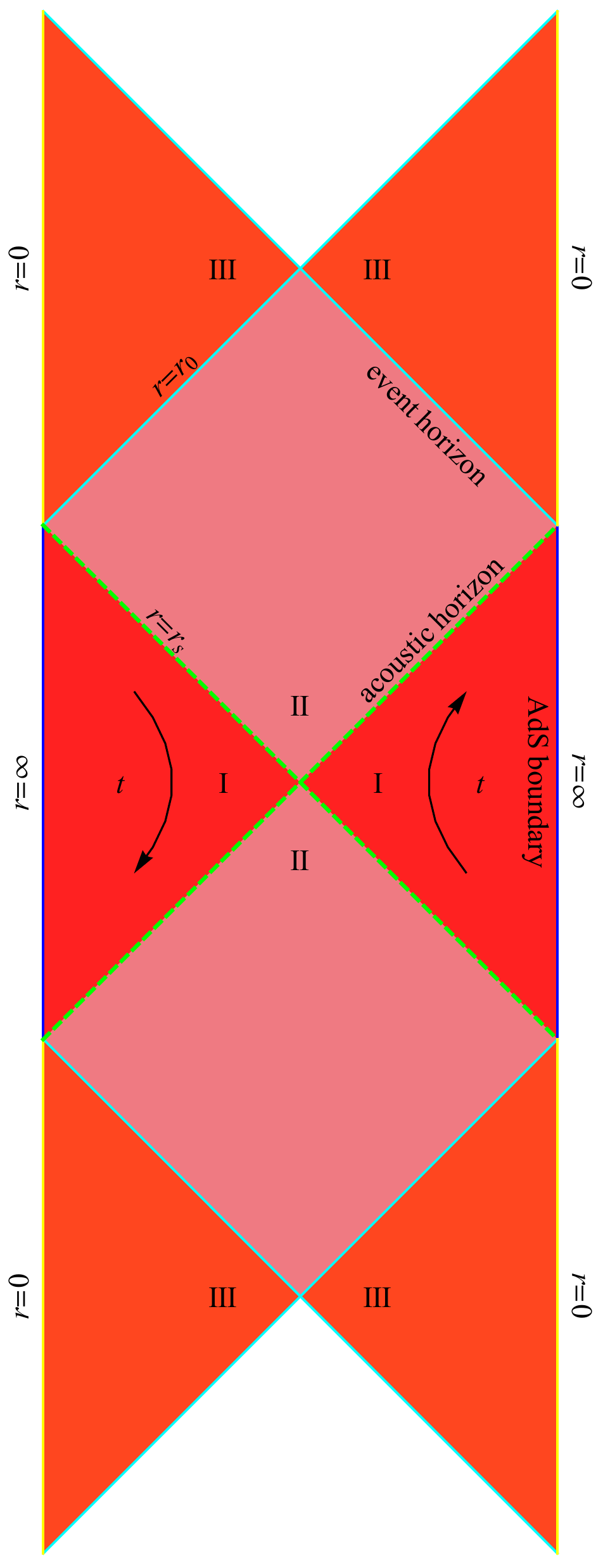}
		\caption{The Penreos diagram of the acoustic AdS black hole \cite{Zanelli}.
			The left and right blue lines and yellow lines respectively correspond to the AdS boundary and singularity. The dotted green lines correspond to the acoustic horizon, and the upper and lower oblique cyan lines stand for the event horizon.}
		\label{fig:PenroseFigure}
	\end{figure}

	Now we construct a thermofield double state (TFD) by exploring the Penrose diagram of acoustic black holes (see Appendix A for the definition of TFD). The schematic diagram is shown in Figure \ref{fig:PenroseFigure}, in which \uppercase\expandafter{\romannumeral1} denotes the region outside the acoustic horizon, \uppercase\expandafter{\romannumeral2} the region between the acoustic horizon and the event horizon, and \uppercase\expandafter{\romannumeral3} the region inside the event horizon.
	The degree of entanglement between subregions $A$ and $B$ on the boundary of the AdS space-time can be quantified by the entropy $S_{A \cup B}$ or the mutual information $I(A;B)$, where the mutual information between subregions $A$ and $B$ is defined as $I(A;B)=S_A+S_B-S_{A \cup B}$. The entropy $S_{A \cup B}$ represents the entropy of the whole system composed of $A$ and $B$. The larger the entropy difference between the sum of the entropy of regions $A$ and $B$ and the entropy of the whole system, the higher the entanglement degree of the two regions.
	That is the greater the mutual information, the greater the entanglement in regions $A$ and $B$. To obtain the scrambling time in this context, one can do the following. At the initial time $t=0$, assuming that the subsystems of the region $A$ and region $B$ are maximally entangled. Then we can turn on a perturbation in the AdS space-time on the left that evolves over time and eventually affects the whole system. After the perturbation decays, the mutual information between two subsystems drops to zero, which means that the entanglement disappears. The time cost in this process corresponds to the scrambling time.

	The purpose of this paper is to study the mutual information $I$ between regions $A$ and $B$ with unconnected boundaries in the maximally extended eternal acoustic-AdS space-time. In the simplest case of the largest extended Penrose diagram of the acoustic-AdS black hole, which has two causally disconnected boundaries, one can be interpreted as an identical copy of the thermofield double state on the second boundary \cite{Maldacena2001}.
	In \cite{GeXH2019,GeXH2020}, some of us considered the realization of acoustic black holes in curved spacetime. Since such relativistic analogue black holes may be realized in high energy process, it is interesting to examine that how two acoustic black holes are first entangled from the beginning and how their mutual information decays away later.

	This paper is organized as follows. In section \ref{The toy model}, we take the (2+1)-dimensional acoustic black holes embedded in AdS space as a toy model and study the maximum extension of the whole space-time by introducing the Kruskal-Szekeres coordinates. We then consider the effects of shock waves on the bulk spacetime and regard the geodesic distance as a physical quantity to reflect the shock wave impact on the space-time geometry. In section \ref{Mutual information and scrambling time}, we apply the Ryu-Takayanagi formula to calculate the mutual information between the left and right subsystems. The results show that, under a proper physical approximation, when the shock wave duration reaches the order of the scrambling time, the entanglement between subsystems disappears and the mutual information is vanishing. The discussion and the conclusion are given in section \ref{Discussion and conclusion}. We briefly review the TFD in Appendix \ref{AppendixA}.

	\section {The Set up} \label{The toy model}
	
	In this section, we discuss the causal structure of the (2+1)-dimensional acoustic black hole embedded in AdS spacetime. The metric is given in equation \eqref{metric2}. In order to obtain the maximum extension of the two-side geometry, we find that it is useful to employ the Kruskal transformation. We first focus on the geometry in the Kruskal-Szekeres coordinates and then consider shock waves in this background. The geodesic length will also be calculated.
	
	\subsection{Kruskal-Szekeres coordinates}
	To facilitate the calculation of the entanglement degree between the two regions at the space-time boundary about acoustic black holes in AdS, one needs to perform some coordinate transformations. In this subsection, we can consider using Kruskal-Szekeres coordinates to describe acoustic black holes. The tortoise coordinate is defined as
	\be
	r_{*}=\int \frac{dr}{f(r)}
	=\frac{\sqrt{3} }{2 \left(r_s ^2-r_0^2\right)}
	\left(r_0 \ln \frac{r+r_0}{|r-r_0|}
	+ r_s \ln \frac{|r-r_s|}{r+r_s}\right).
	\ee
	Considering that $r_0$ and $r_s$ are basically of the same order of magnitude, we make the following approximation
	\bea
	r_{*}
	&=&\frac{1}		
	{2 \kappa_s}
	\left(\frac{r_0}{r_s} \ln \frac{r+r_0}{r-r_0}
	+ \ln \frac{r-r_s}{r+r_s}\right)\nonumber\\
	&\approx&
	\frac{1}		
	{2 \kappa_s}
	\ln 		
	\frac{\left( r-r_s \right) (r+r_0)}
	{\left( r+r_s \right) (r-r_0)},
	\eea
	where the surface gravity is given by $\kappa_s=2 \pi T=2 \pi/ \beta=f'(r_s)/2=\frac{r_s}{\sqrt{3}}  \left(1-\frac{r_0^2}{r_s^2}\right)$.
	We can further introduce the null light Kruskal coordinates $U$ and $V$
	\be\label{KSC0}
	U=-e^{-\kappa_{s}u},~~~V=e^{\kappa_{s}v},
	\ee
	where $u=t-r^{*}$, $v=t+r^{*}$.

	The Kruskal-Szekeres coordinates are denoted by $(V,U,\phi)$, where $\phi$ is the coordinate same as equation \eqref{metric2}, but $t$ and $r$ are replaced
	by $V$ and $U$ based on the following coordinate transformation:
	\begin{subequations}\label{KSC1}		
		\begin{equation}\label{KSC11}\left.
			\begin{aligned}
				U_R&=-\sqrt{
					\frac{\left( r-r_s \right) (r+r_0)}
					{\left( r+r_s \right) (r-r_0)}}\
				e^{-\kappa_s  t}\\				
				V_R&=
				\sqrt{
					\frac{\left( r-r_s \right) (r+r_0)}
					{\left( r+r_s \right) (r-r_0)}}\
				e^{\kappa_s  t}
			\end{aligned}\right\}r>r_s \ (-+),
		\end{equation}
		\begin{equation}\label{KSC12}\left.
			\begin{aligned}
				U_L&=
				\sqrt{
					\frac{\left( r-r_s \right) (r+r_0)}
					{\left( r+r_s \right) (r-r_0)}}\
				e^{-\kappa_s  t}\\				
				V_L&=-
				\sqrt{
					\frac{\left( r-r_s \right) (r+r_0)}
					{\left( r+r_s \right) (r-r_0)}}\
				e^{\kappa_s  t}
			\end{aligned}\right\}r>r_s \ (+-).
		\end{equation}
	\end{subequations}
	
\begin{figure}[htbp]
		\centering
		\includegraphics[scale=0.6]{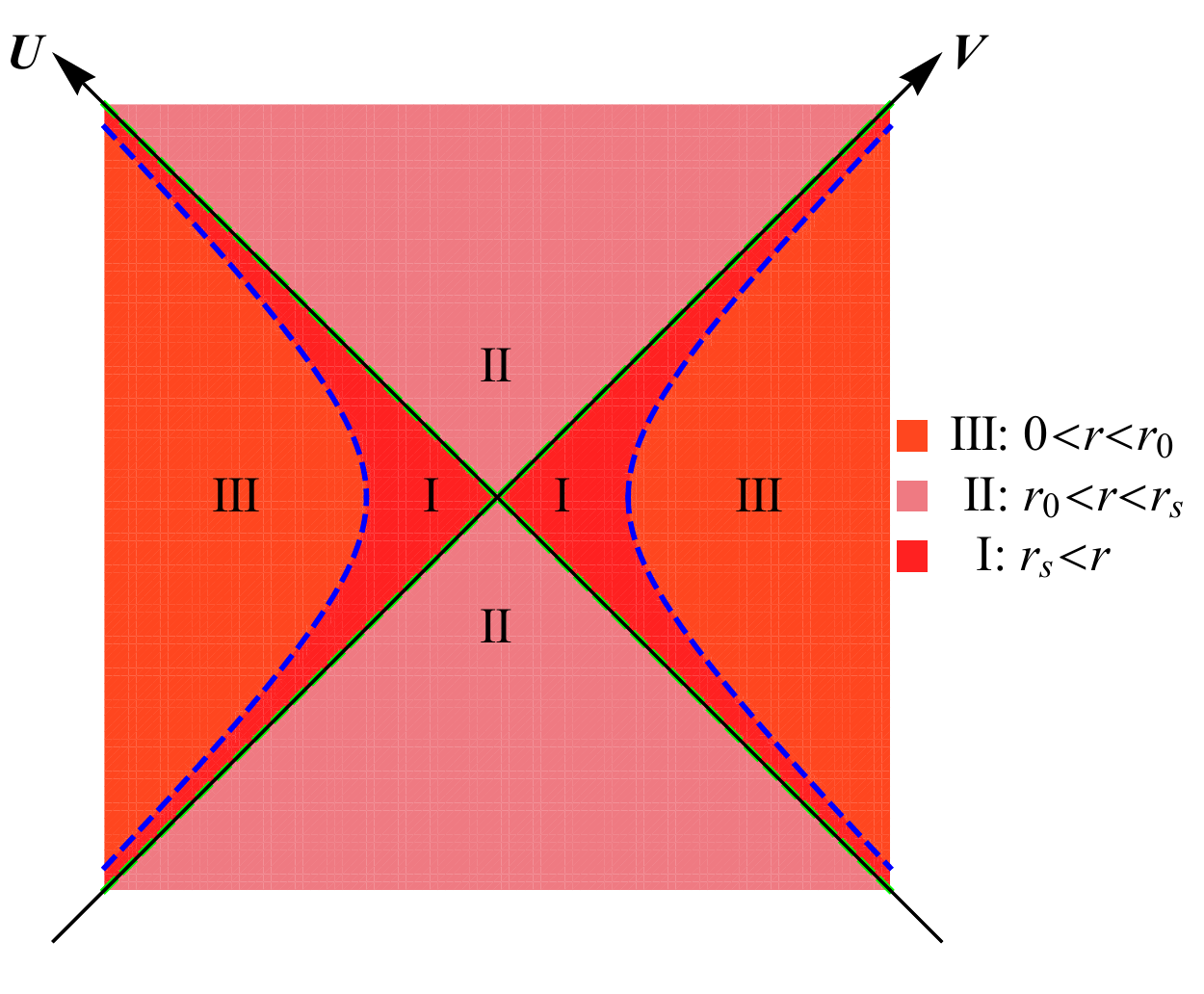}
		\caption{The $U$-$V$ plane of the Kruskal extension: The region \uppercase\expandafter{\romannumeral1}, \uppercase\expandafter{\romannumeral2}, \uppercase\expandafter{\romannumeral3} corresponds to region $r>r_s$, $r_0<r<r_s$ and $0<r<r_0$ respectively. The green solid line represents the acoustic horizon $r=r_s$. The blue dotted line represents the location $r=0$ and $r=\infty$. The location $r=r_0$ is at infinity. We can see that the isochronal line $t=0$ is a horizontal line through the origin. For $t>0$, an isochron is a straight line sloping upward from the left region to the right region. When $t \rightarrow \infty $, the isochronal line is a line approaching the $V$ axis. For $t<0$, the situation is just the opposite.}
		\label{fig:KruskalFigure}
	\end{figure}

	In terms of the Kruskal-Szekers coordinate, the metric given in \eqref{metric2} can be recast as
	\begin{equation}\label{KSC2}
		ds^2=
		-
		\frac{\sqrt{3} r_s^2 (r-r_0)^2 (r+r_s)^2}{r^2 \left(r_s^2-r_0^2\right)^2}
		dUdV
		+r^4 d\phi^2\!,
	\end{equation}
	where $r$ is considered as a function $r=r(V,U)$, satisfying the following relation	
	\begin{equation}\label{KSC3}
		 V U  = - \frac{\left( r-r_s \right) (r+r_0)}
		 {\left( r+r_s \right) (r-r_0)} ,		
	\end{equation}
	which can be derived from equation (\ref{KSC1}). The Kruskal-Szekeres metric (\ref{KSC2}) does not has singularity at $r=r_s$ and $r=r_0$. When $r_s=r_0$, this metric has singularity, but not physically allowed. This indicates that the singularity in \eqref{metric2} is just a coordinate singularity rather than a physical singularity.
	From (\ref{KSC3}), we can see that lines of constant $r$ are hyperbolas of constant $U V$ in the $U\mbox{-}V$ plane. Similarly, isochrones can be expressed as curves in the Kruskal diagram (Figure \ref{fig:KruskalFigure}). From (\ref{KSC0}) and \eqref{KSC1}, one can obtain
	\begin{equation}\label{KSC51}
			\frac{V}{U}=-e^{2 \kappa_s t}.
	\end{equation}
    Thus lines of constant $V/U$ are lines of constant $t$-straight lines through the origin.	
	From \eqref{KSC2}, we can see that sound cones correspond $d\phi=0$ (radial) and $ds^2=0$ (null). There are curves of sound cones
	\begin{equation}
		V=\text{const},\  U=\text{const}
	\end{equation}
	in Figure \ref{fig:KruskalFigure}.
	The acoustic horizon at $r=r_s$ is the line with $V=0$ or $U=0$.
	Replacing coordinates $U$ and $V$ in \eqref{KSC3} with two new coordinates $U'$ and $V'$, we have	
		\begin{align}
			V'&=\tan^{-1}V,\\
			U'&=\tan^{-1}U.
		\end{align}
	The Penrose diagram is shown in the $U'V'$ plane as demonstrated in Figure \ref{fig:PenroseFigure} (see also \cite{Zanelli}).

	\subsection{Acoustic shock waves}
	Assuming that the total system is in its thermal equilibrium at temperature $T$, now we have a thermofield double state (TFD) of the two CFTs. For a review of the TFD, one may refer to Appendix A. We are interested in how the shock waves affect the TFD and how spacetime structure will be modified.
 A simply way to turn on the shock waves can be realized adding a few quasiparticles say ``phonons" at the boundary and leting them fall into the acoustic black hole.

\begin{figure}[htbp]
		\begin{minipage}[c]{0.5\textwidth}
			\centering
			\includegraphics[scale=0.5]{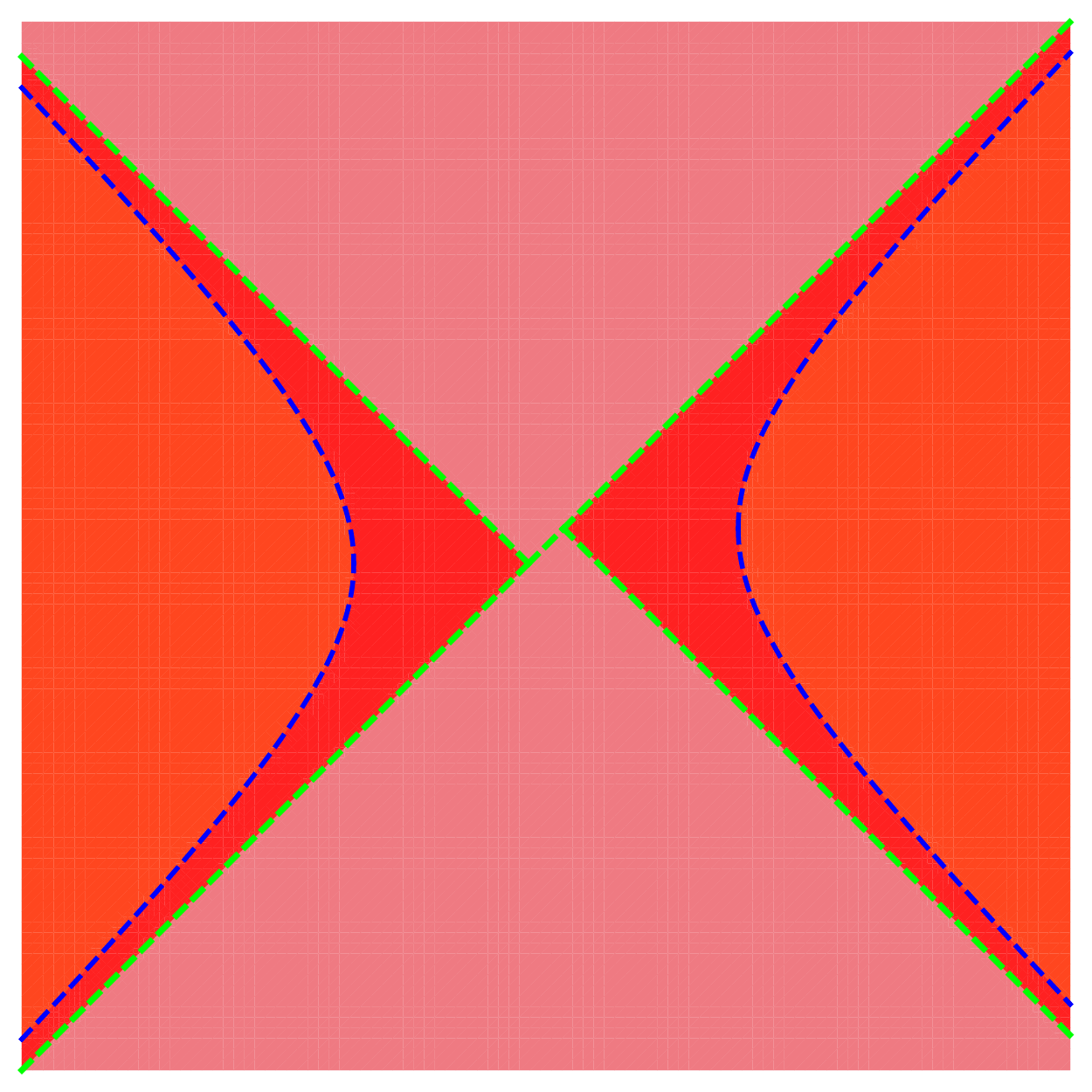}
		\end{minipage}
		\begin{minipage}[c]{0.5\textwidth}
			\centering
			\includegraphics[scale=0.4]{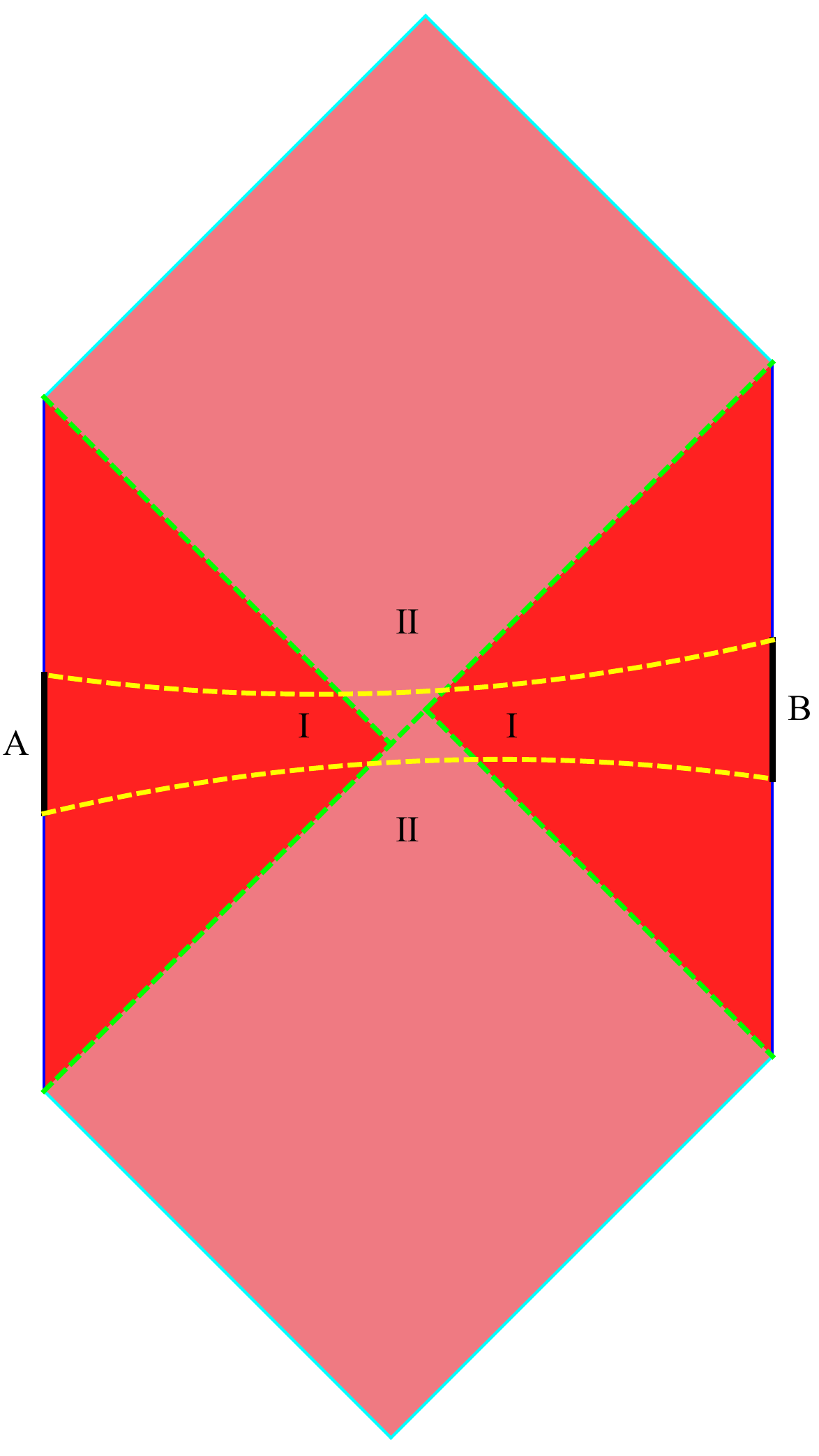}
		\end{minipage}
\caption{ The Kruskal-Szekeres 	and the Penrose schematic diagrams for the geometry with a shock wave from the left boundary. The green line represents the acoustic horizon $r=r_s$. The blue line in the Kruskal figure represents the location of $r=0$ and $r=\infty$. We can see the green lines with $UV=0$ and acoustic horizons is staggered by a displacement $\alpha$. The regions $A$ and $B$ are in the AdS boundary, which is black line sketched in Penrose diagram and the length of the black line corresponds to the distance $d^{\prime}$ in formula \protect\eqref{distance2}. The distance $d$ in formula \protect\eqref{distance1} corresponds to the yellow dotted line in the figure.
}		\label{fig:ShockWaveFigure}
	\end{figure}

    We release a quasiparticle with energy $E_0$ from the boundary, at time $t_p$ in the past, so at time $t=0$ it has energy $E_P$
	\be
	E_P \sim E_0 e^{\kappa_s  t_p}.
	\ee
	If the time $t_p $ is long enough, then the energy $E_P$ will be large enough that we need to consider its reaction to space-time geometry. We consider gluing geometry of mass $M$ $(M \propto r_s^2)$ to geometry of mass $M+E_0$ through the null surface with $U_p=e^{-\kappa_s t_p}$ \cite{Stephen}. Here $E_0$ is the asymptotic energy of perturbation, which is small compared to $M$.
	Considered the metric after the perturbation, as demonstrated in Figure \ref{fig:ShockWaveFigure}, the right and the left of the region \uppercase\expandafter{\romannumeral1} are chosen with coordinates $U,V$ and $ \widetilde{U},\widetilde{V}$ respectively.
	The shell $U_p=e^{-\kappa_s  t_p}, 
	\widetilde{U}_p=e^{-\widetilde\kappa_s  t_p}$ propagates at the infinity.
	For small $E_0$, we approximate $U_p \simeq \widetilde{U}_p$. The matching condition for $\widetilde{V}$ to $V$ via $\widetilde{U}_p \widetilde{V}=-e^{-\widetilde\kappa_s  \widetilde r_*}, U_p V=-e^{-\kappa_s r_*}$.
	Because $U V=-\frac{\left( r-r_s  \right) (r+r_0)}
	{\left( r+r_s  \right) (r-r_0)} ,V / U =-e^{2 \kappa_s  t}$,
	we have \cite{Stephen}
	\be
	\frac{\left( r-r_s  \right) (r+r_0)}
	{\left( r+r_s  \right) (r-r_0)}
	=			
	\left( r-r_s  \right)	
	C \left( r  \right),
	\ee
	where $C$ is smooth at the acoustic horizon $r=r_s$ and $C(r_s)\neq 0$. We obtain
	\be
	C \left( r \right)	
	=
	\frac{ r+r_0 }
	{\left( r+r_s  \right) (r-r_0)}.
	\ee
	According to \cite{Stephen}, for linear order in $E_0$ and large $t_p$, we find the shift of the shock wave to the geometry $\alpha=\widetilde{V}-V$ and obtain
	\bea
	\alpha
	&=&
	\frac{E_0}{U_p}\frac{dr_s }{dM}C(r_s)\nonumber\\
	&=&
	\frac{E_0}{U_p}\frac{dr_s }{dM}
	\frac{ r_s +r_0 }{2 r_s  (r_s -r_0)}.
	\eea
	Considering the entropy of acoustic black hole $S=\pi r_s/2$, $dS=\pi /2 \ dr_s$ \footnote{The entropy of acoustic black hole is considered as contributed by the generalized entropy of the Hawking radiation of the quasiparticles (phonons).}, we then have $dM=T dS
	=\frac{\kappa_s}{4} dr_s$ and $dr_s/dM=	4/\kappa_s$. Thus the shift of spacetime metric is
	\bea
	\alpha	
	&=&
	\frac{E_0}{U_p}\frac{dr_s }{dM}
	\frac{ r_s +r_0 }{2 r_s  (r_s -r_0)}
	\nonumber\\	
	&=&
	\frac{4 E_0 }{\kappa_s }
	\frac{ r_s +r_0 }{2 r_s  (r_s -r_0)}
	e^{\kappa_s  t}
	\nonumber\\	
	&=&
	\frac{2 \sqrt{3}	E_0}{  \left( r_s -r_0 \right)^2}
	e^{\kappa_s  t}.
	\eea
	Using discontinuous coordinates $\mathcal{U}=U$, $\mathcal{V}=V+\alpha \theta(U)$, we can obtain a shock wave geometry with the form
	\be
	ds^2=
	-
	\frac{\sqrt{3} r_s^2 (r-r_0)^2 (r+r_s)^2}{r^2 \left(r_s^2-r_0^2\right)^2}
	[d\mathcal{U}d\mathcal{V}+\alpha \delta(\mathcal{U})d\mathcal{U}^2]
	+r^4 d\phi^2\!.
	\ee

	\subsection{Geodesics}
	For the purpose of observing the effect of the perturbation on the entanglement entropy between subsystems, we need to calculate the entanglement entropy through the RT formula. According to the RT formula, the entanglement entropy between a given space-like subregion $A$ and its complementary in a $d$-dimensional CFT is given by the smallest surface area homologous to $A$ in the dual $(d+1)$-dimensional bulk spacetime \cite{Ryu}. While for the case of $(2+1)$-dimensional spacetime, the entanglement entropy of two regions $A$ and $B$ respectively in left and right AdS spacetime is related to its geodesic distance.
	
      For computing the geodesic distances $d$ between points in the AdS geometry, we can use the distance formula for the geodesic distance in pure AdS$_{2+1}$
	\be\label{distance}
	\cosh d = T_1 T_1' + T_2 T_2' - X_1 X_1'-X_2X_2',
	\ee
	On the right hand, the embedded coordinates are defined as follows
		\begin{align}			
			T_1 &= \frac{V+U}{1+UV}
			= \frac{ \sqrt{\left(r^2-r_s ^2\right) (r^2-r_0^2) } \sinh \kappa_s   t}{r \left( r_s- r_0\right)} ,
			\notag\\		
			T_2 &= \frac{1-UV}{1+UV}\cosh \kappa_s  \phi
			= \frac{\left(r_s  r_0- r^2\right) \cosh \kappa_s  \phi}{r \left(r_s- r_0\right)},
			\notag\\		
			X_1 &= \frac{V-U}{1+UV}
			= \frac{ \sqrt{\left(r^2-r_s ^2\right) (r^2-r_0^2) } \cosh \kappa_s   t}{r \left( r_s- r_0\right)},
			\notag \\		
			X_2 &= \frac{1-UV}{1+UV}\sinh \kappa_s  \phi
			= \frac{\left(r_s  r_0- r^2\right) \sinh \kappa_s  \phi}{r \left(r_s- r_0\right)} .
			\label{embedded}		
		\end{align}	
	The coordinates $T_1, T_2, X_1, X_2$ satisfy the following relationship
	\be
	T_1^2+T_2^2-X_1^2-X_2^2 \equiv 1.
	\ee
	In the left asymptotic region, one need to add the imaginary part $i\beta/2$ (i.e. $i\pi/\kappa_s$) to time $t$ in \eqref{embedded}, which means that $U$ and $V$ in the Kruskal coordinates have extra minus sign (i.e. use formula \eqref{KSC12} to replace formula \eqref{KSC11} in formula \eqref{embedded}).
	
	Taking $T_1'=X_1'=V$, $T_2'=\cosh \kappa_s  \phi$, $X_2'=\sinh \kappa_s  \phi$ in \eqref{distance}, we obtain the geodesic distance $d_2$ from right boundary to the acoustic event horizon. That is
	\be
	\cosh d_2=
	\frac{ e^{-\kappa_s  t_R} v \sqrt{\left(r^2-r_s ^2\right) (r^2-r_0^2) }-r^2+r_0 r_s}{r (r_0-r_s)}.
	\ee
	Similarly, the geodesic distance $d_1$ from left boundary to the acoustic event horizon is
	\be
	\cosh d_1=
	\frac{e^{-\kappa_s  t_L} (\alpha +v) \sqrt{\left(r^2-r_s ^2\right) (r^2-r_0^2) }+r^2-r_0 r_s}{r (r_s-r_0)}.
	\ee
	The total geodesic distance $d$ between two points on different boundaries is
	\be
	d
	=
	2 \log \left(\frac{\alpha  e^{-\frac{1}{2} \kappa_s  (t_L+t_R)} \sqrt{\left(r^2-r_0^2\right) \left(r^2-r_s^2\right)}+\left(r^2-r_0 r_s\right) \cosh \frac{\kappa_s  (t_L-t_R)}{2}}{r (r_s-r_0)}\right).
	\ee
	In the large $r$ limit, we approximate
	\bea \label{distance1}
	d
	&\approx&
	2 \log \left(\frac{\alpha  e^{-\frac{1}{2} k (t_L+t_R)} r^2+2 r^2 \cosh \left(\frac{1}{2} k (t_L-t_R)\right)}{ r (r_s-r_0)}\right) \notag \\
	&=&2 \log \left(\frac{\alpha  e^{-\frac{1}{2} k (t_L+t_R)} r+2 r \cosh \left(\frac{1}{2} k (t_L-t_R)\right)}{  (r_s-r_0)}\right)\notag \\
	&=&2 \log \left(\frac{2 r}{r_s-r_0}\right)
	+
	2 \log \left(\frac{\alpha  e^{-\frac{1}{2} k (t_L+t_R)}} {2}+ \cosh \frac{\kappa_s  (t_L-t_R)}{2}\right).
	\eea
	The above situation corresponds to the distance from $A$ to $B$ in Figure \ref{fig:ShockWaveFigure} the yellow dotted line. Similarly, on the same boundary, the geodesic distance $d^{\prime}$ between two equal-time points $(r,t,\phi_1)$ and $(r,t,\phi_2)$ is
	\be \label{distance2}
	d^{\prime}\approx
	2 \log \left(\frac{2 r \sinh \frac{\kappa_s  (\phi_1 -\phi_2)}{2}   }{r_s-r_0}\right)
	=
	2 \log \left(\frac{2 r }{r_s-r_0}\right)
	+2 \log \left(\sinh \frac{\kappa_s  (\phi_1 -\phi_2)}{2} \right).
	\ee
	Because in 2 + 1-dimensional geometry, the regions on the AdS boundary is one-dimensional. At the same time, the extremum surface becomes a line. This distance corresponds to the length of $A$ or $B$ (the black line in Figure \ref{fig:ShockWaveFigure}).
	
	\section{Mutual information and scrambling time}  \label{Mutual information and scrambling time}
	Up to now, we have constructed the TFD through a toy model and obtained geodesic distances \eqref{distance1} and \eqref{distance2} in the geometry after applying the perturbation. Now the primary goal is to obtain the mutual information $I(A;B)$ between the subsystems $A$ and $B$ (see Figure \ref{fig:ShockWaveFigure}). We first focus on the entanglement entropy of the subsystem $S_A$, or equivalently $S_B$.  From the expression \eqref{distance1} and \eqref{distance2}, there are two suitable extemal surfaces (distance). However, the RT formual tells us that the smallest surface is the most proper choice. Therefore, we chose \eqref{distance2} to obtain the smaller value when $\phi<\pi$, i.e.
	\bea
	S_A=S_B &=&\text{Min}\bigg[\frac{\text{Area}(\gamma)}{4G_N}\bigg] \notag \\
             &=&\frac{d^{\prime}}{4G_N}=
	\frac{1}{2G_N}
	\left[
	\log \left(\frac{2 r }{r_s-r_0}\right)
	+\log \left(\sinh \frac{\kappa_s  (\phi_1 -\phi_2)}{2}\right)
	\right].
	\eea
	Note that the position of the area $A$ and $B$ we selected are the same on their respective boundaries.
	Secondly, consider the entanglement entropy of the total system $S_{A \cup B}$. The first choice of the total entropy is to add the two region's entanglement entropy $S_{A \cup B}^{(1)}=S_A+S_B$, giving the result as $I(A;B)=\big[ S_A+S_B\big]-S_{A \cup B}^{(1)} \equiv 0$, which is trivial. But for the second choice, using \eqref{distance1} under the equal-time condition $t_L=t_R$, we connect a pair of geodesic from $A$ to $B$, which lead to the minimal value
	\bea
	S_{A \cup B}^{(2)}&=& 2 \times \frac{d}{4G_N} \\ \notag
	&=&\frac{1}{G_N}
	\left[
	\log \left(\frac{2 r}{r_s-r_0}\right)
	+
	\log \left(\frac{\alpha  } {2}+ 1\right)
	\right].
	\eea
	Thus the mutual information is obtained as
	\be\label{mutual information}
	I(A;B)=
	S_A+S_B-S_{A \cup B}^{(2)}\\
	=
	\frac{1}{ G_N}	
	\left[
	\log \left(\sinh \frac{\kappa_s  (\phi_1 -\phi_2)}{2}\right)
	-
	\log \left(\frac{\alpha  } {2}+ 1\right)	
	\right].
	\ee
	So, when $\sinh \frac{\kappa_s  (\phi_1 -\phi_2)}{2} < 1$, i.e. a small regions with $A$ and $B$, there is $S_{A \cup B}^{(2)}>S_{A \cup B}^{(1)}$, which means $I(A;B)=0$ for any value of $\alpha$.
	But for larger regions and a small $\alpha$, $S_{A \cup B}^{(2)}$ is smaller than $S_{A \cup B}^{(1)}$, and the expression \eqref{mutual information} represents the evolution of the entanglement between the left and right subsystems with perturbation in the thermofield double state. The perturbations eventually affect the whole spacetime, resulting in complete purification of the entanglement between subsystems. When the mutual information is vanishing, we have
	\be
	\sinh \frac{\kappa_s  (\phi_1 -\phi_2)}{2}
	=
	\frac{\alpha  } {2}+ 1.
	\ee
	Invoking the following approximation
	\be
	\sinh \frac{\kappa_s  (\phi_1 -\phi_2)}{2}
	\approx
	\frac{1}{2}
	e^{\frac{\kappa_s  (\phi_1 -\phi_2)}{2}},
	\ee
	and assuming that $\alpha \gg 1$, we have
	\be
	\frac{1}{2}
	e^{\frac{\kappa_s  (\phi_1 -\phi_2)}{2}}
	=
	\frac{\sqrt{3}	E_0}{  \left( r_s -r_0 \right)^2}
	e^{\kappa_s  t}.
	\ee
	Finally, we obtain
	\be\label{Scr}
	t
	=
	\frac{ \phi_1 -\phi_2}{2}
	+\frac{1}{\kappa_s }\ln \frac{ 2 \sqrt{3} E_0}{  \left( r_s-r_0 \right)^2}.
	\ee
	In the above expression, we have taken the entropy of acoustic black hole with following equation
	\be
	\frac{2 S}{\beta}
	=
	\frac{\left( r_s+r_0 \right)\left( r_s-r_0 \right)}{ 2 \sqrt{3}}.
	\ee
	In the limit $r_s \gg r_0$, the scrambling time is obtained as
	\be\label{scrambling}
	t_w
	\approx
	\frac{1}{\kappa_s }\ln \frac{2 S}{\beta E_0}
	\sim
	\frac{\beta}{2\pi}\ln S,
	\ee
	where $E_0 \sim T=1/ \beta$.  The results are consistent with the Hayden-Preskill experiment considered in \cite{Hayden2007,Stephen}.
	In other words, after the perturbation, the time for the system to return to the equilibrium state is directly proportional to the logarithm of the entropy of the acoustic black hole. The scrambling time is inversely proportional to the temperature, that is, the lower the temperature, the longer the scrambling time. This is consistent with the existing thermodynamic experience. Similar analyses can yield a similar result with $\phi>\pi$, the specific details can be found in \cite{Stephen}.

	\section{Discussion and conclusion}  \label{Discussion and conclusion}
	 In summary, we study the time evolution of the entanglement entropy and the mutual information of subsystems in the thermofield double state under the perturbation and use acoustic black holes embedded in AdS space to simulate this process. We first construct the TFD by embedding acoustic black holes into the AdS spacetime \eqref{metric2}. Next, we calculate the geodesic distance \eqref{distance1} and \eqref{distance2} in the whole black hole spacetime after the Kruskal extension.
	 The last result \eqref{scrambling} shows the influence of the shock wave on the scrambling time. The correlation between the two regions $A$ and $B$ in the AdS boundary of the TFD decays exponentially with time after the shock wave.
	 It's shows the entanglement between dual spacetime is destroyed, and this is related to the entropy of the central acoustic black hole. After adding shock waves, the more information the central acoustic black hole contains, the longer the entanglement of dual space-time can last.

     For this result \eqref{Scr}, we can also see that the order of the scrambling time is related to the size of the acoustic horizon and the optical horizon.
     If we substitute $\kappa_s$ into equation \eqref{Scr}, then there is $ t\propto \frac{\sqrt{3} r_s}{r_s^2-r_0^2 }\ln \frac{ 2 \sqrt{3} E_0}{  \left( r_s-r_0 \right)^2}$.
     Within the physical allowable range, the larger the acoustic horizon $r_s$, the smaller the scrambling time. The size of the real event horizon $r_0$ has the opposite effect on the scrambling time. If the real event horizon $r_0$ increases, the scrambling time will be longer.
     We think this is so interesting.
     As for the relationship between $t$ and $r_0$, we may simply understand that the negative temperature $\beta$ at the acoustic horizon is positively correlated with $r_0$ ($\beta=\frac{2\sqrt{3} \pi r_s}{r_s^2 -r_0^2}$). That is in formula \eqref{scrambling}, the larger $r_0$, the larger $\beta$, so the final scrambling time will be longer.
     Although the entropy at the acoustic horizon $S$ is related to $r_s$ is positively correlated ($S=\frac{\pi r_s}{2}$), but $\beta$ is negatively correlated with $r_s$, so in the final result of formula \eqref{scrambling}, the larger $r_s$ is, the smaller the scrambling time is.
     It requires further experimental verification.

      On the other hand, we find that there is a similar scrambling time for (2+1)-dimensional acoustic black holes, which provides some theoretical information for simulating black hole entanglement and black hole information loss problem in subsequent acoustic experiments.
	Of course, some discrete models may need to be considered in the experiment, which needs defining the relevant experimental design in the future. For acoustic black holes in AdS spacetime, the scrambling time have a negative correlation with the radius of the acoustic horizon and a positive correlation with the real event horizon. In this regard, this paper provides a theoretical result for the expectation of subsequent experiments. Therefore, as an attempt and exploration, we considered whether there are similar possible effects in table-top acoustic black holes to the black hole information problem, so as to provide some hints for subsequent theoretical and experimental studies.
	At the last, our toy model is similar to the rotating BTZ in metric factor (in formula \eqref{metric2}), and we recently find some studies about shock waves in the rotating BTZ in \cite{Malvimat}. Our main result \eqref{scrambling} is similar with the result formula (4.17)  in this paper ($t_*=\frac{\beta(1-\mu)}{\pi}\log \frac{S}{\beta E_{eff}}$). Because form of the metric factor for acoustic balck holes embedded in AdS space, this may provide a new way and idea for laboratory simulation of rotating BTZ.
	This also inspires us to consider simulating rotating BTZ in the form of acoustic black holes in the future.

	\section*{Acknowledgements}
	We would like to thank Yu-Qi Lei, Cheng-Yuan Lu and Hai-Ming Yuan for helpful discussions. This work is partly supported by NSFC, China (Grant No.11875184 $\&$ No.11805117).

	\renewcommand{\thesection}{\Alph{section}}
	\setcounter{section}{0}

	\section{Thermofield double state}  \label{AppendixA}
In this appendix, we review the thermofield double state. The TFD can be prepared by using two identical copies of a quantum mechanical system. That is to say, preparing two identical copies of a quantum system described by Hamiltonians $H_1=H_2$, with eigenstates $| n\rangle_1$  and $|n\rangle_2$ of a common eigenenergy  $E_n$.  A thermofield double state with standard representation is defined as
\be
|TFD\rangle=\frac{1}{\sqrt{Z}}\sum_{n}e^{-\frac{\beta E_n}{2}} |n\rangle_{L} \otimes |n\rangle_{R}.
\ee
Notice that the relation $(H_{L}-H_{R})|TFD\rangle=0$ is satisfied for TFD and these two systems are maximally entangled.

We consider an operator $V$ at the time $t=0$ and let the system evolves until time $t$, we impose the operator $W_L$ on the $L$-system. The physical state becomes
\be
|\psi(t)\rangle=W_{L}(t) |TFD\rangle.
\ee
After that, we evolve the state inversely. The two-point correlator function of the $L$ and $R$ systems is then given by
\be
F^{OTO}_{\beta}(t)=\frac{1}{Z}{\rm tr}\bigg(e^{-\frac{\beta H_{L}}{2}}W^{\dag}_{L}(t)V^{\dag}_L(0)W_{L}(t)e^{-\frac{\beta H_{L}}{2}}V_{L}(0)\bigg).
\ee


\begin{thebibliography}{10}
		\bibitem{Abbott2016}
		B. P. Abbott, T. Adams, R. Bonnand, et al. (LIGO Scientific Collaboration and Virgo Collaboration)  GW151226: Observation of gravitational waves from a 22-solar-mass binary black hole coalescence, Physical Review Letters, 116, 241103 (2016).
		
		\bibitem{Abbott2017}
		B. P. Abbott, R. Abbott, T. D. Abbott, et al.  GW170817: Observation of gravitational waves from a binary neutron star inspiral, Physical Review Letters, 119, 161101 (2017).
		
		\bibitem{EHT2019}
		The Event Horizon Telescope Collaboration, First M87 Event Horizon Telescope Results. I. The Shadow of the Supermassive Black Hole, Astrophys. J. 875, L1 (2019).
		
		\bibitem{Nova}
		J. R. Nova,  K. Golubkov,   V. I. Kolobov,  J. Steinhauer,    Observation of thermal hawking radiation and its temperature in an analogue black hole, Nature. 569, 688 (2019).
		
		\bibitem{Cromb}
		M. Cromb,   G. M. Gibson,  E. Toninelli,  M. J. Padgett,  E. M. Wright,  D. Faccio,  Amplification of waves from a rotating body, Nature Physics, 1  (2020).
		
		\bibitem{Nation}
		P. D. Nation,  M. P. Blencowe, A. J. Rimberg,  E. Buks,   Analogue hawking radiation in a dc-squid array transmission line, Physical Review Letters, 103(8), 087004 (2009).
		
		\bibitem{Giovanazzi}
		S. Giovanazzi,  Hawking radiation in sonic black holes, Physical Review Letters, 94(6), 061302 (2004).
		
		\bibitem{Jacobson}
		T. A. Jacobson, G. E. Volovik, Event horizons and ergoregions in 3He, Physical Review D, 58(6) (1998).
		
		\bibitem{Garay}
		L. J. Garay, J. R. Anglin,  J. I. Cirac,  P. Zoller,   Sonic black holes in dilute bose-einstein condensates, Physical Review A, 63(2), 184  (2000).
		
		\bibitem{Hu2019}
		J. Hu, L. Feng, Z. Zhang, C. Cheng, Quantum simulation of unruh radiation, Nature Physics, 15(8), 1 (2019).
		
		\bibitem{Horstmann}
		B. Horstmann, B. Reznik, S. Fagnocchi, J. I. Cirac, Hawking radiation from an acoustic black hole on an ion ring, Physical Review Letters, 104, 250403  (2010).
		
		\bibitem{Philbin}
		T. G. Philbin, C. Kuklewicz,  S. Robertson, S. Hill, F. Konig, U. Leonhardt, Fiber-optical analog of the event horizon, Science, 319 (2008).
		
		\bibitem{Unruh}
		W. G. Unruh, Experimental black-hole evaporation, Physical Review Letters, 46(21), 1351  (1981).
		
		\bibitem{Drori2019}
		J. Drori, Y. Rosenberg, D. Bermudez, Y. Silberberg, U. Leonhardt, Observation of Stimulated Hawking Radiation in an Optical Analogue, Physical Review Letters, 122, 010404 (2019).
		
		\bibitem{Sheng2016}
		C. Sheng, H. Liu, S. Zhu, D. A. Genov, Omnidirectional optical attractor in structured gap-surface plasmon waveguide, Scientific Reports, 6, 23514 (2016).
		
		\bibitem{Sheng2013}
		C. Sheng, H. Liu, Y. Wang, et al. Trapping light by mimicking gravitational lensing, Nature Photon, 7, 902  (2013).
		
		\bibitem{Cheng2010}
		Q. Cheng, T. J. Cui, W. X. Jiang, B. G. Cai, An omnidirectional electromagnetic absorber made of metamaterials, New Journal of Physics, 12, 063006 (2010).
		
		\bibitem{Koke2016}
		C. Koke, C. Noh, D. G. Angelakis, Dirac equation in 2-dimensional curved spacetime, particle creation, and coupled waveguide arrays, Annals of Physics, 374, 162  (2016).
		
		\bibitem{Wang2020}
		Y. Wang, C. Sheng, Y. Lu, J. Gao, Y. Chang, X. Pang, T. Yang, S.	Zhu, H. Liu, X. Jin, Quantum simulation of particle pair creation near the event horizon, National Science Review, 7(9), 1476  (2020).
		
		\bibitem{Zehua}
        Z. Tian, Y. Lin, U. R. Fischer, J. Du, Testing the upper bound on the speed of scrambling with an analogue of Hawking radiation using trapped ions, The European Physical Journal C, 82, 212 (2022).
		
		\bibitem{Smolyaninov2014}
		I. I. Smolyaninov, Holographic duality in nonlinear hyperbolic metamaterials, Journal of Optics, 16: 075101 (2014).
		
		\bibitem{Li1}
		L. Miao, R. X. Miao, P. Yi, More studies on metamaterials mimicking de sitter space, Optics Express, 18(9), 9026  (2009).
		
		\bibitem{Li2}
		M. Li, R. X. Miao, Y. Pang, Casimir energy, holographic dark energy and electromagnetic metamaterial mimicking de sitter, Physics Letters B, 689(2-3), 55  (2010).
		
		\bibitem{Smolyaninov2010}
		I. I. Smolyaninov, E. E. Narimanov, Metric signature transitions in optical metamaterials, Physical Review Letters, 105(6), 067402 (2010).
		
		\bibitem{Sekino2008}
		Y. Sekino, L. Susskind, Fast scramblers, Journal of High Energy Physics, 2008(10) (2008).
		
		\bibitem{Page1993}
		D. N. Page, Average Entropy of a Subsystem, Physical Review Letters, 71, 1291 (1993).
		
		\bibitem{Hayden2007}
		P. Hayden, J. Preskill, Black holes as mirrors: quantum information in random subsystems, Journal of High Energy Physics, 2007(9), 887  (2007).
					
		\bibitem{GeXH2019}
		X. H. Ge, M. Nakahara, S. J. Sin, Y. Tian, S. F. Wu, Acoustic black holes in curved spacetime and the emergence of analogue minkowski spacetime, Physical Review D, 99(10), 104047 (2019).
		
		\bibitem{GeXH2020}
		Q. B. Wang, X. H. Ge, Geometry outside of acoustic black holes in $(2+1)$-dimensional spacetime, Physical Review D, 102(10), 104009 (2020).
		
		\bibitem{Visser}
		M. Visser, Acoustic black holes: horizons, ergospheres, and hawking radiation, Classical and Quantum Gravity, 15(6), 1767 (25) (1997).
		
		\bibitem{Maldacena2001}
		J. Maldacena, Eternal black holes in anti-de sitter, Journal of High Energy Physics, 2003 (2003).
		
		\bibitem{Zanelli}
		M. Banados, M. Henneaux, C. Teitelboim, J. Zanelli, Geometry of the $(2+1)$-black hole, Physical Review D, 48(4), 1506  (1993).
		
		\bibitem{Stephen}
		S. H. Shenker, D. Stanford, Black holes and the butterfly effect, Journal of High Energy Physics, 2014, 67 (2014).
		
		\bibitem{Ryu}
		S. Ryu, T. Takayanagi, Holographic derivation of entanglement entropy from AdS/CFT, Physical Review Letters, 96, 181602 (2006).
		
		\bibitem{Malvimat}
		V. Malvimat, R. R. Poojary, Fast scrambling due to rotating shockwaves in BTZ, (2021).   [arXiv:2112.14089 [hep-th]]
		
	\end{thebibliography}
\end{document}